\documentclass[preprint,amsmath,amssymb,amssymb,aps,prl]{revtex4}
\usepackage[colorinlistoftodos, color=green!40, prependcaption]{todonotes}

\usepackage{graphicx,url}
\usepackage[utf8]{inputenc}  
\usepackage[english]{babel}
\usepackage{amsmath}
\usepackage{amssymb}
\usepackage{physics}
\usepackage{float}
\usepackage{url}
\usepackage{soul}
\usepackage{graphicx}
\usepackage{stmaryrd}
\usepackage{aeguill}
\usepackage{dcolumn}
\usepackage{bm}
\usepackage{braket}
\usepackage{times}
\usepackage{mathrsfs}
\usepackage{color}
\usepackage{todonotes}
\usepackage[pdftex, pdftitle={Article}, pdfauthor={Author}]{hyperref}
\usepackage{natbib}
\begin{document}

\title{Interlayer Exciton Valleytronics in Bilayer Heterostructures Interfaced with a Metasurface}

\author{Mandar Sohoni$^{1}$, Pankaj K. Jha$^{2}$, Muralidhar Nalabothula$^{3}$, Anshuman Kumar$^{1,}$\footnote{anshuman.kumar@iitb.ac.in}}
\affiliation{$^{1}$Physics Department, Indian Institute of Technology Bombay, Mumbai 400076, India\\
$^{2}$Department of Applied Physics and Materials Science, California Institute of Technology Pasadena, California 91125, $^{3}$Physics and Materials Science Research Unit, University of Luxembourg, 162a avenue de la Faïencerie, L-1511 Luxembourg, Luxembourg}

\begin{abstract}
The recent proposal of using an anisotropic vacuum for generating valley coherence in transition metal dichalcogenide (TMDC) monolayers has expanded the potential of such valley degrees of freedom for applications in valleytronics. In this work, we open up a completely new regime, inaccessible with monolayer TMDCs, of  spontaneously generated valley coherence in interlayer excitons in commensurate TMDC bilayer heterostructures. Using the peculiar out of plane polarization of interlayer excitons in conjunction with an in-plane anisotropic electromagnetic vacuum, we show that a much larger region of the Bloch sphere can be accessible in these heterostructures. We show the accessible phases of these excitons given this in-plane anisotropic electromagnetic vacuum. Our analysis of spontaneous coherence for interlayer excitons may pave the way for engineering an array of interacting quantum emitters in Moir\'e heterostructures.
\end{abstract}
\date{\today}
\maketitle
\noindent
In recent years, valleytronics -- the technology to manipulate the electronic `valley degree of freedom' in two dimensional gapped Dirac systems, which possess pairs of degenerate band extrema or valleys, has received enormous attention for information processing applications \cite{vtron,Langer2018}. Few layer transition metal dichalcogenides (TMDCs) are one class of semiconductors that show great promise for such technologies. TMDCs are van der Waals materials, which means that their heterostructures can host excitons in which the electron and hole are confined to separate layers, i.e., interlayer excitons \cite{Miller2017, Arora2017}. In TMDC heterostructures, such as $\text{MoX}_{2}/\text{WX}_{2}$, there exists a type II band alignment, which results in the interlayer exciton being the most energetically favorable \cite{Rivera2018}. In addition, their heterostructures are rather easily formed without the constraints of epitaxial matching, and they can be integrated with other photonic platforms \cite{Liu2019,Novoselov2016}. There have been several proposals to use TMDC heterostructures to develop novel optoelectronic devices, similar to those in the field of spintronics \cite{Cao2012, magento}. In this work, we open up a new regime of spontaneous valley coherence for interlayer excitons in commensurate TMDC bilayer heterostructures, exhibiting extremely rich coherence features. Recently, it was shown that interlayer excitons in bilayer TMDC heterostructures not only couple to in-plane polarized light, but also to $z$ polarized light due to the spatial separation of the electron hole pair \cite{selrules, r0exist}. 
We have found that this results in the existence of orthogonal dipoles that are not in-plane circularly polarized in such systems.
\\
\\
It is of interest to study spontaneous valley coherence generation for interlayer excitons for the following reasons. Firstly, interlayer excitons, due to the spatial separation of the electron-hole pair, have longer radiative lifetimes as compared to their intralayer counterparts \cite{intexcliftm}, as well as much smaller intervalley scattering rates \cite{intexcscat}. Secondly, moire superlattices can generate an array of such interlayer excitons and thus, have the potential to host interacting quantum emitter arrays \cite{moire,Seyler2019,Jin2019,PhysRevB.97.035306}. Lastly, due to the form of the valley exciton coupling in intralayer excitons, circularly polarized dipoles will only have a real coupling constant for diagonal Green's tensors which limits the state vector to only reside in the $xz$ plane of the Bloch sphere. Interlayer excitons on the other hand, while retaining in-plane circularly polarized dipoles, also have out of plane orthogonal dipoles for certain interlayer translations. These non-trivial dipoles have complex coupling constants, thus resulting in a larger accessible region of the Bloch sphere.
\\
\\
As with any scheme for quantum control, one must be able to generate coherence between these two excitons. Broadly, there are two approaches to generate such coherence between the valleys -- one that involves an external coherent electromagnetic pump \cite{extfieldcoh,Hao2016,Zhu2014} and another, spontaneously, using an anisotropic vacuum generation \cite{bp,svc}. To generate the anisotropic vacuum we used a custom designed array of nano-antennas, i.e., a metasurface. Metasurfaces, a special kind of planar metamaterial, allow user defined electromagnetic waveform responses with control over properties such as amplitude, phase, and polarization \cite{Jahani2016, met2, met3} with an ultra-thin spatial footprint. Figure 1 shows a schematic of the proposed TMDC heterostructure placed above the metasurface.
\\
\\
Our letter is organized as follows. Firstly we look at orthogonal interlayer exciton dipole moments in commensurate stackings of bilayer TMDC heterostructures, based on the analysis done in \cite{selrules}, for singlet and triplet interlayer excitons, followed by identifying the interlayer translation for the non-trivial orthogonal dipole moments in $\pm$K valleys. Secondly we analyze the effect of an anisotropic electromagnetic vacuum on these orthogonal dipoles. Lastly we propose a metasurface and bilayer $\text{MoSe}_{2}/\text{WSe}_{2}$ heterostructure to observe the coherence between the two valleys, and metrics such as the Stokes' parameters to quantify the coherence.
\\
\\
Valley excitons ($+K$ or $-K$) in monolayer TMDCs have excitons that are circularly polarized in the plane of the monolayer \cite{TMDCbasics}. Such systems be modelled as a V-level scheme with the two transitions being orthogonal, but degenerate. Valley excitons in TMDC bilayer heterostructures, however, have excitons with dipole moments that can couple to right circularly, left circularly, and linearly ($z$) polarized light \cite{selrules}. In general, the interlayer valley exciton dipole moments can be written as
\begin{equation}
\mathbf{D}_{1} = a^{+K}_{+}\mathbf{e}_{+} + a^{+K}_{-}\mathbf{e}_{-} + a^{+K}_{z}\mathbf{e}_{z}, \quad
\mathbf{D}_{2} = a^{-K}_{+}\mathbf{e}_{+} + a^{-K}_{-}\mathbf{e}_{-} + a^{-K}_{z}\mathbf{e}_{z}
\end{equation}
where $\mathbf{D}_{1} = \mathbf{D}_{+K\pm K}$ and $\mathbf{D}_{2} = \mathbf{D}_{-K\mp K}$, depending on the stacking (R/H). The coefficients $a^{\pm K}_{\pm,z}$ are dependent on the interlayer translation $\bar{r}_{0}$ (see supplementary information). In this work, we are interested in the forms of the orthogonal dipoles, i.e., when $\hat{\mathbf{D}}_{2}^{*}\cdot\hat{\mathbf{D}}_{1} = 0$, because these dipoles can then be viewed as a pseudo-spin.
\\
\\
Most of the values of $r_{0}$ for which $\hat{\mathbf{D}}_{2}^{*}\cdot\hat{\mathbf{D}}_{1} = 0$ are the high symmetry points ($r_{0} = \frac{1}{3}$ or $\frac{2}{3}$) where the dipoles are circularly polarized in the $xy$ plane, similar to the monolayer case. However, there are a few `non-trivial' zeros where the orthogonal dipoles are not in-plane circularly polarized, but have all three components in the labarotary frame. It turns out that these non-trivial zeros are in fact circularly polarized dipoles (due to time reversal symmetry), up to a phase, with a quantization axis that is not the $z$ axis. The quantization axis $\bar{n}$ of these non-trivial zeros is given by
\begin{equation}
\bar{n}_{1(2)} = \text{Re}(\hat{\mathbf{D}}_{1(2)})\times\text{Im}(\hat{\mathbf{D}}_{1(2)})
\end{equation}
\noindent
and since $\hat{\mathbf{D}}_{2}^{*}\cdot\hat{\mathbf{D}}_{1} = 0 \Rightarrow \text{Re}(\hat{\mathbf{D}}_{1(2)}).\text{Im}(\hat{\mathbf{D}}_{1(2)}) = 0$, the non-trivial dipoles are circularly polarized in the plane defined $\bar{n}_{1(2)}$ \cite{compvec} (note that $\bar{n}_{1} = -\bar{n}_{2}$, see supplementary information).
\\
\\
Even though commensurate TMDC bilayer heterostructures with most of the interlayer translation values, $r_{0}$, do not exist naturally, they can be found in incommensurate heterostructures with small twist angles (moire superlattices) \cite{r0exist, intcpl}. They are found on length scales larger than the monolayer lattice constants, but smaller than the moire supercell lattice vectors. In these regions, the atom placement is indistinguishable from that of a commensurate stacking with interlayer translation. Such excitons are therefore experimentally realizable. Figure 2 (b) shows the variation of $|\hat{\mathbf{D}}_{2}^{*}\cdot\hat{\mathbf{D}}_{1}|$ with $r_{0}$ for singlet excitons for the example of a commensurate H type stacking, which is the focus of this work. Our formalism can be easily extended to other types of stacking, which are further discussed in the supplementary information.
\\
\\
In an isotropic electromagnetic vacuum, the coupling between any two orthogonal transitions is forbidden. One proposal to work around this was to create an anisotropic vacuum that would result in a non-zero coupling \cite{anivacprop}. Anisotropic vacuums can be generated by using metasurfaces \cite{supmir}, or by using anisotropic polaritonic materials \cite{bp}. We briefly elucidate how an anisotropic vacuum can enable coupling in the following. An emission process in one valley (say $+K$, in monolayer TMDCs) will result in a photon that will not be able to excite the other valley in free space. Once the vacuum becomes anisotropic, the interaction term between the vacuum field created by the excited dipole and the dipole of the other valley, i.e., $\bar{\mathbf{d}}_{-K}^{*}\cdot\bar{\mathbf{E}}_{+K}$, is non-zero. The coupling rate between the two orthogonal dipoles is given by \cite{imG}
\begin{equation}
\Tilde{\kappa}_{21} = \frac{2\omega_{0}^{2}}{\hbar\epsilon_{0}c^{2}}\mathbf{D}_{2}^{*}\cdot\text{Im}\big[\overset{\leftrightarrow}{\mathbf{G}}\big]\cdot\mathbf{D}_{1}
\end{equation}
\noindent
where $\overset{\leftrightarrow}{\mathbf{G}}$ is the total Green's tensor at the position of the dipoles (Note that $\text{Re}\big[\overset{\leftrightarrow}{\mathbf{G}}\big]$ is typically small for nano-antenna metasurfaces). The normalized coupling rate, $\Tilde{\kappa}_{21}/\gamma_{0}$ ($\gamma_{0}$ is the free space decay rate), for in-plane circular dipoles evaluates to $\frac{\gamma_{xx} - \gamma_{yy}}{2}$, where $\gamma_{aa}$ is the normalized decay rate for a dipole with $\hat{a}$ polarization, i.e., $\gamma_{aa} \propto \mathbf{a}^{*}\cdot\text{Im}\big[\overset{\leftrightarrow}{\mathbf{G}}\big]\cdot\mathbf{a}$. In an anisotropic vacuum, where $\gamma_{xx}\neq\gamma_{yy}$, the coupling will be a finite non-zero value. The normalized coupling constant for our case can be written as $\kappa = \Tilde{\kappa}_{21}/\gamma_{0} = \kappa_{xx} + \kappa_{yy} + \kappa_{zz}$, where 
\begin{equation}
\kappa_{xx,yy} = \pm\frac{\gamma_{xx,yy}\big(a^{+K}_{+} \pm a^{+K}_{-}\big)^{2}}{2\big(|a^{+K}_{+}|^{2} + |a^{+K}_{-}|^{2} + |a^{+K}_{z}|^{2}\big)},
\end{equation}
\begin{equation}
\kappa_{zz} = \frac{\gamma_{zz}\big(a^{+K}_{z}\big)^{2}}{|a^{+K}_{+}|^{2} + |a^{+K}_{-}|^{2} + |a^{+K}_{z}|^{2}}.
\end{equation}
\noindent
If $\gamma_{xx} = \gamma_{yy} = \gamma_{zz}$, the coupling constant will be proportional to the dot product of the two dipoles and will be 0 for orthogonal transitions. A table of the values of $\kappa/\gamma$ for different stackings and types (R and H; singlet and triplet) for the non-trivial zeros can be found in the supplementary information. In order to show how the intervalley coupling rate varies as a function of interlayer translation, in Figure 2 (a) we plot the variation of the imaginary and real parts of $\kappa/\gamma$ for the singlet exciton  in a commensurate H type $\text{MoSe}_{2}/\text{WSe}_{2}$ heterostructure. In Figure 2 (a), for a fair comparison between the different interlayer translation, we consider a metasurface which is optimized such that $\gamma_{xx} = 0.1$, and $\gamma_{yy} = \gamma_{zz} = 1$ for each value of the translation. The values of $\gamma_{ii}$ chosen are in accordance with what was expected from a metasurface with a large numerical aperture. Figure 2 (c) shows the variation of the same parameters with $\gamma_{xx}$ ($\gamma_{yy,zz} = 1$) for the exciton marked in Figure 2 (b), which confirms that our chosen values of the  $\gamma_{ii}$'s are indeed close to optimal for observing the largest coupling rate. It must be noted however, that in the case of other types of stacking, the optimal values of the corresponding decay rates might differ.
\\
\\
In this work we use a nano-antenna based metasurface to generate an anisotropy between the $x$ and $y$ directions. Our metasurface is designed to act as a normal mirror for $y$ polarized light and a spherical mirror for $x$ polarized light. To mimic such a phase profile, the dimensions of each nanoantenna and its position on the metasurface have been designed accordingly. This will result in destructive interference between the emitted and reflected fields for $x$ polarized light at the dipoles' position while leaving the field of $y$ polarized light unaffected, hence causing a suppression of the radiative rate in the $x$ direction. Figure 3 (a) shows the phase profile for $x$ polarized light that we are trying to mimic. Figure 3 (b) shows the phase of the reflected $x$ polarized light for each of our chosen nano-antennas. Our designed metasurface gives us $\gamma_{xx} \approx 0.1$ and $\gamma_{yy} = 1$, which is quite close to the optimal value as highlighted in Figure 2 (c).
\\
\\
To measure the coherence between the two dipoles, a quantity known as the Degree of Linear Polarization (DoLP) has been proposed \cite{Dolp}. This is the normalized Stokes' parameter $\text{S}_{1}$ \cite{stokes}. For systems such as ours, this metric only captures a part of the coherence in the system, since it corresponds to a projective measurement in the $xz$ plane of the Bloch sphere. To capture the full coherence, we also need to consider the third Stokes' parameter, $\text{S}_{2}$. In order to calculate these experimentally measurable parameters, we solve for the density matrix of the system via the master equation given below \cite{oscME}:
\begin{equation}
\dot{\rho} =
i\sum_{n}\Delta\omega_{n}\big[\ket{n}\bra{n}, \rho\big] + 
i\sum_{m \neq n}\delta_{nm}\big[\ket{n}\bra{m}, \rho\big] + 
\sum_{n\neq m}\Gamma_{nm}\bigg(\rho_{mn}\ket{g}\bra{g} - \frac{1}{2}\big\{\ket{n}\bra{m}, \rho\big\}\bigg)
\end{equation}
\begin{equation}
+ \sum_{n}\Gamma_{nn}\bigg(\rho_{nn}\ket{g}\bra{g} - \frac{1}{2}\big\{\ket{n}\bra{n}, \rho\big\}\bigg)
\end{equation}
\noindent
where $n,m = 1,2$. $\ket{g}$ denotes the ground state, $\Gamma_{21} = \kappa = \Gamma_{12}^{*}$ is the coupling constant and captures the two photon (virtual) process between the two orthogonal dipoles, $\Gamma_{nn}$ is the spontaneous decay rate of the dipole, $\delta_{nm}$ is the resonant dipole-dipole interaction, and $\Delta\omega_{n}$ accounts for the lamb-shift. For our system, both $\delta_{nm}$ and $\Delta\omega_{n}$ are two orders of magnitude smaller than $\Gamma_{nm}$ and have hence been neglected.
\\
\\
To propose a method to experimentally measure the coherence, we excite the system with a bi-directional incoherent pump for both $\hat{\mathbf{D}}_{1}$ and $\hat{\mathbf{D}}_{2}$. In steady state, the Stokes' parameters take the form (see supplementary information for the rate equations)
\begin{equation}
\frac{\text{S}_{2}}{\text{S}_{0}} = \frac{i(\rho_{12} - \rho_{21})}{\rho_{11} + \rho_{22}} = \frac{\kappa_{i}}{\beta}
\end{equation}
\begin{equation}
\frac{\text{S}_{1}}{\text{S}_{0}} = \frac{\rho_{12} + \rho_{21}}{\rho_{11} + \rho_{22}} = -\frac{\kappa_{r}}{\beta}
\end{equation}
\noindent
where $\kappa_{r(i)}$ is the real (imaginary) part of $\kappa$, $\gamma$ is the spontaneous decay rate of the dipoles in the presence of the metasurface, $\gamma_{s}$ is the intervalley scattering rate,  $R$ is the incoherent pump rate, and $\beta = \gamma + R + \gamma_s$. Figure 4 (a) shows the evolution of the Stokes parameters on the Poincar\'e sphere with bi-directional incoherent pumping for both $\hat{\mathbf{D}}_{1}$ and $\hat{\mathbf{D}}_{2}$ with the initial state prepared in $\hat{\mathbf{D}}_{1}$. For our heterostrucutre, the system evolves outside the $xz$ plane. For comparison, we have included a trajectory for a system where $\text{Im}[\kappa] = 0$. Figure 4 (b) shows the steady state value of the Stokes' parameters against the intervalley scattering rate. $S_{1}$ and $S_{2}$ contain information on the final state of the system on the Bloch sphere. A non-zero value of $S_{2}$ indicates that the system has left $xz$ plane of the Bloch sphere. In Figure 4 (c), we show that with metasurfaces of different design, the phase $\phi = \tan^{-1}{(S_{2}/S_{1})}$ of the projection of the polarization state on the equatorial plane can be tuned for different values of $\gamma_{xx}$ and $\gamma_{yy}$ ($\gamma_{zz} = 1$). One should note that $\gamma_{xx/yy}$ is tunable via the geometrical design of the metasurface and enables accessing different states on the Poincar\'e sphere.
\\
\\
To summarize, we have shown a valleytronics application of spontaneous coherence between interlayer excitons in commensurate TMDC heterostructures, with a study on the phase ($\phi = \tan^{-1}{(S_{2}/S_{1})}$) of the steady state with bi-directional incoherent pumping for both valleys. We have found the existence of `non-trivial' orthogonal dipoles in R and H commensurate stackings of $\text{MoSe}_{2}/\text{WSe}_{2}$ heterostructures. These dipoles, due to time reversal symmetry, are circularly polarized with their quantization axis differing from the $z$ axis, and could potentially be used to observe non-inverse dynamics (different temporal evolutions for different initial states of the emitter) \cite{noninv}, which could lead to the development of simple quantum gates in such systems. We have analyzed the coupling between these kinds of dipoles in the presence of an anisotropic vacuum created by a nano-antenna based metasurface. The couplings are complex, which allow the system to evolve outside the $xz$ plane of the Bloch sphere. For our example system we have considered a singlet exciton in a commensurate  $\text{MoSe}_{2}/\text{WSe}_{2}$ H type heterostructure and have analyzed the population dynamics, and the temporal evolution of our proposed measurable metrics. However, our analysis remains valid for other kinds of stacking in other bilayer systems supporting interlayer excitons. This work paves the way for quantum vacuum engineered heterostructure system, i.e., a Moir\'e lattice that could potentially host an array of interacting quantum emitters.

\emph{Acknowledgement--} AK acknowledges funding support from the Department of Science and Technology via grant numbers SB/S2/RJN-110/2017, DST/NM/NS-2018/49 and SEED grant from Industrial Research and Consultancy Centre, IIT Bombay. 

\newpage
\begin{figure}[H]
    \includegraphics[width = \linewidth]{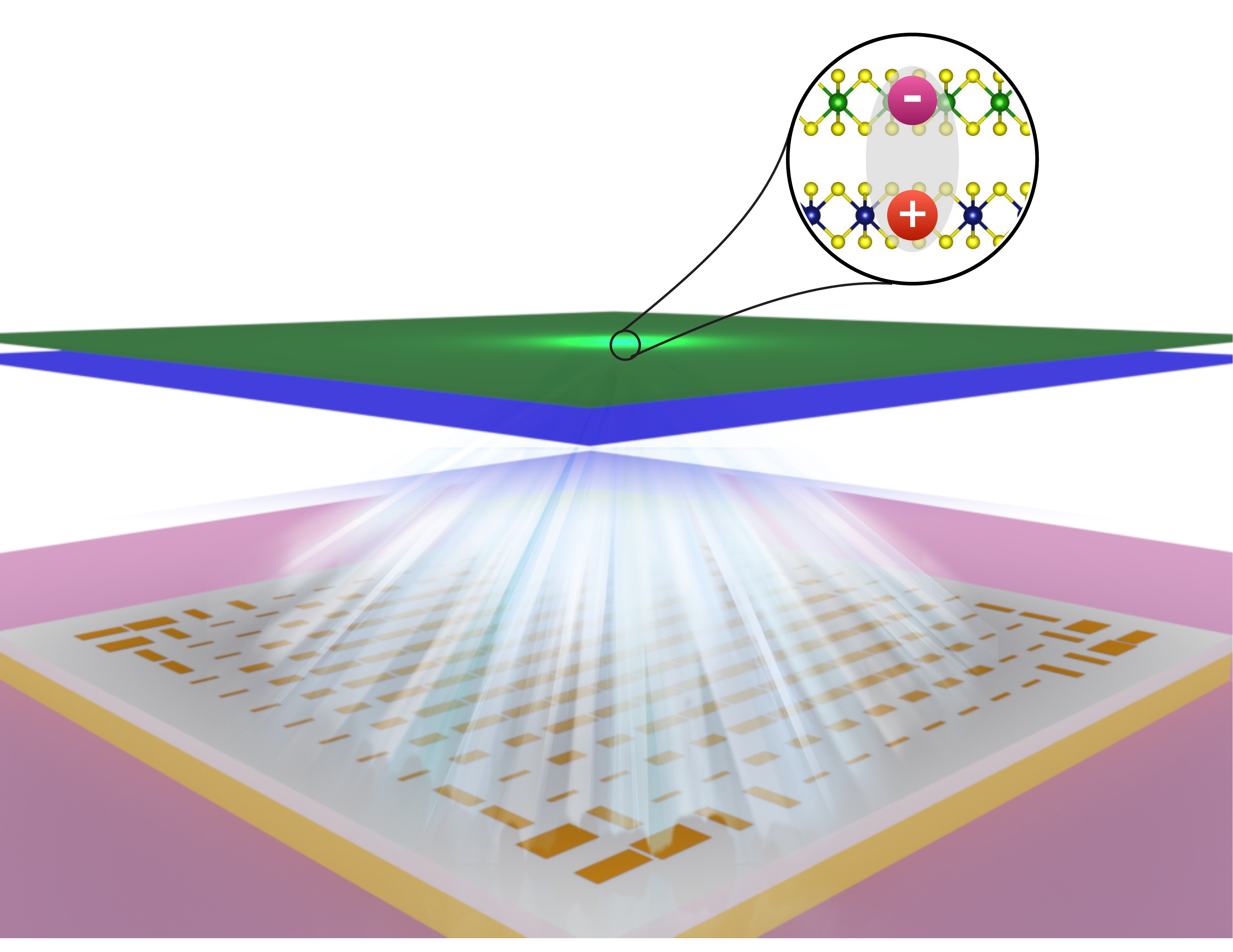}
    \caption{A schematic of a TMDC heterostructure interfaced with a metasurface. The zoomed-in inset depicts the interlayer exciton. In an isotropic electromagnetic vacuum, such as the case with free space, orthogonal interlayer excitons ($\pm K\pm K$ for R type stacking and $\pm K\mp K$ for H type stacking) will not interact with each other. The metasurface creates an anisotropy in the decay rates for $x$ and $y$ polarized dipoles which allows a finite non-zero coupling between orthogonal interlayer excitons.} 
\end{figure}
\newpage
\begin{figure}[H]
    \includegraphics[width = \linewidth]{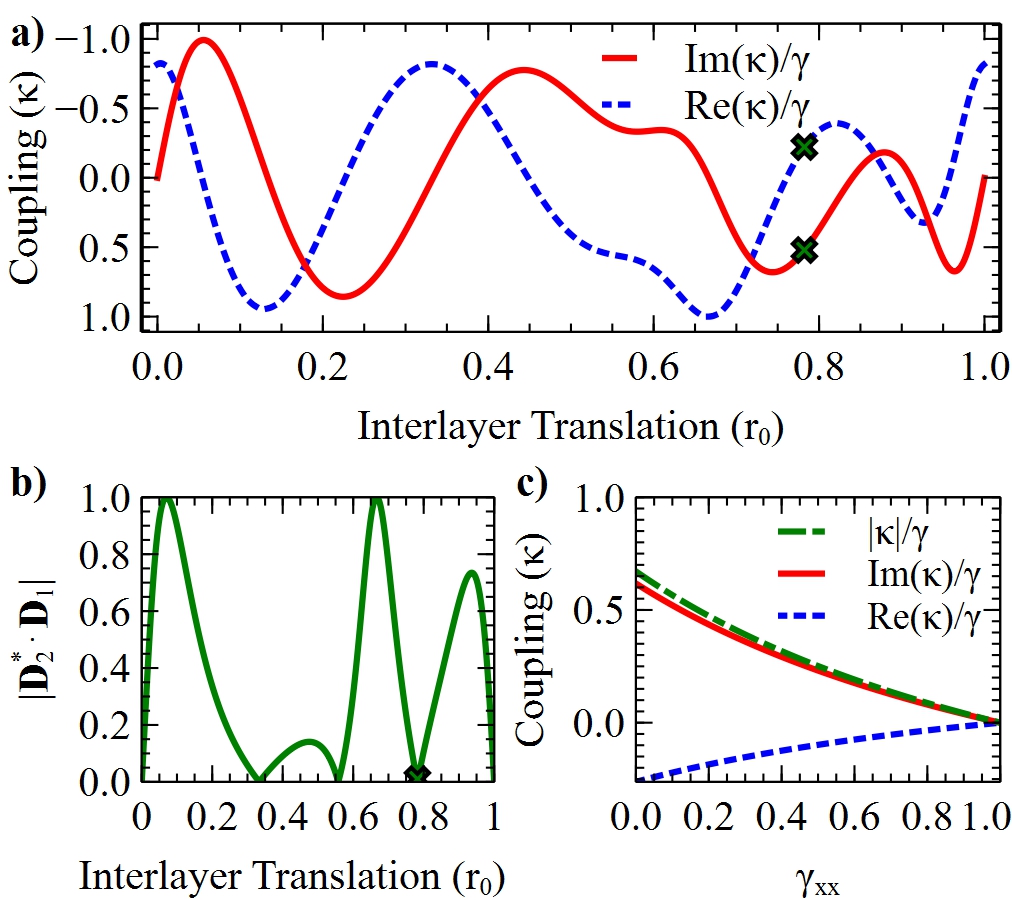}
    \caption{Singlet exciton in a commensurate H type $\text{MoSe}_{2}/\text{WSe}_{2}$ heterostructure : (a) Variation of $\text{Re}(\kappa)/\gamma$ and $\text{Im}(\kappa)/\gamma$ with the interlayer translation $r_{0}$ when $\gamma_{xx} = 0.1$, and $\gamma_{yy} = \gamma_{zz} = 1$, in accordance with the values expected from our metasurface design. The cross marks the values of the real and imaginary part of the coupling at the non-trivial zero marked in (b). (b) The absolute value of the dot product between $\hat{\mathbf{D}}_{2}$ and $\hat{\mathbf{D}}_{1}$ as a function of the interlayer translation $r_{0}$ . The non-trivial zero that we chose to design the metasurface for is marked with a cross. (c) Variation of $\text{Re}(\kappa)/\gamma$, $\text{Im}(\kappa)/\gamma$, and $|\kappa|/\gamma$ with $\gamma_{xx}$ for the dipole marked by the cross in (b). We have taken $\gamma_{yy}$ and $\gamma_{zz}$ to be 1.}
\end{figure}
\newpage
\begin{figure}[h]
    \includegraphics[width = \linewidth]{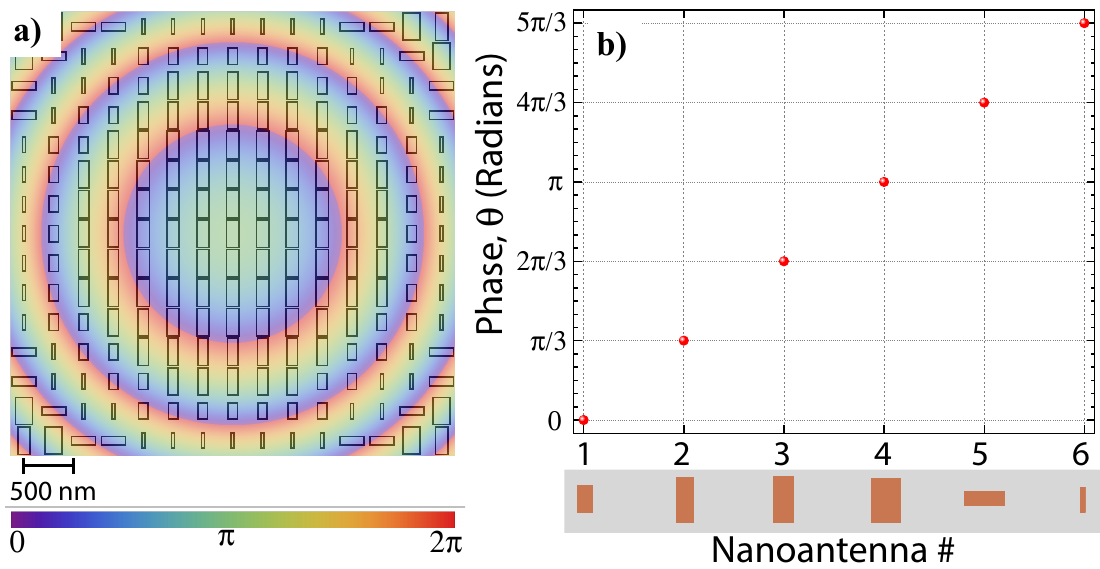}
    \caption{Phase profile of the designed metasurface. (a) The phase profile for the reflected $x$ polarized light from the metasurface. This will result in destructive interference between the emitted and reflected fields for $x$ polarized light at the dipoles' position. (b) The phase of each of our chosen antennas. We have chosen phase values of $0^{\circ}$, $60^{\circ}$, $120^{\circ}$, $180^{\circ}$, $240^{\circ}$, and $300^{\circ}$.} 
\end{figure}
\newpage
\begin{figure}[H]
    \includegraphics[width = \linewidth]{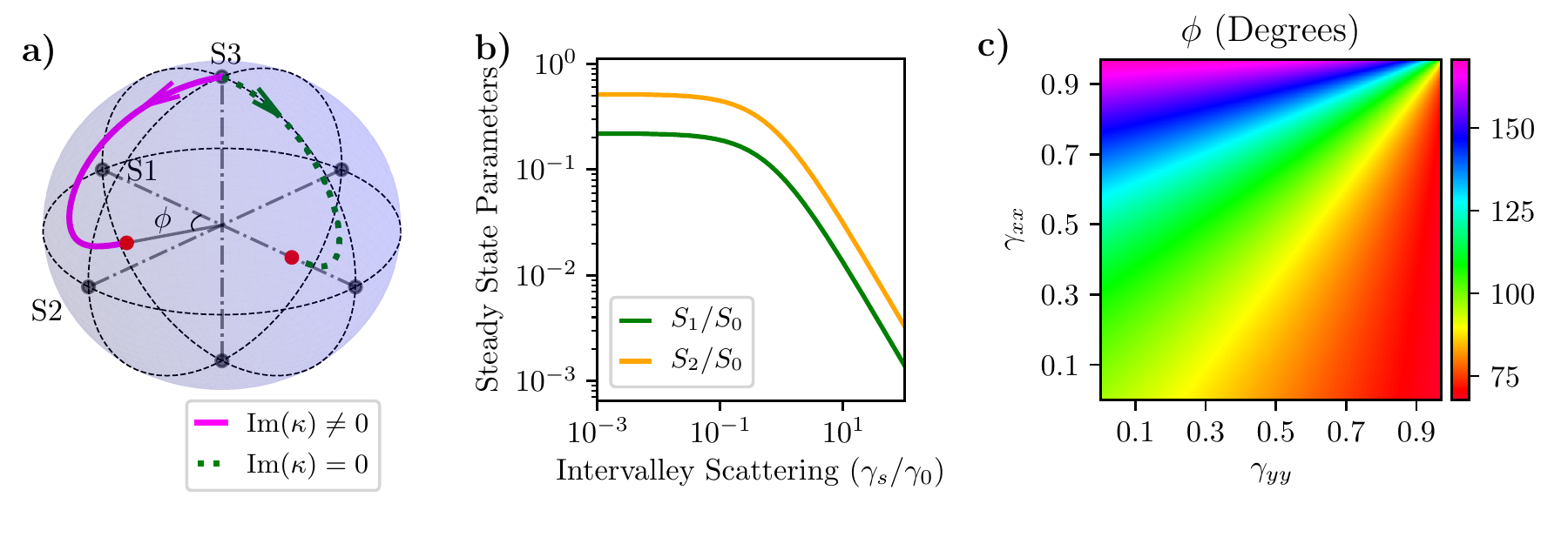}
    \caption{Stokes Parameters with bi-directional incoherent pumping for both $\hat{\mathbf{D}}_{1}$ and $\hat{\mathbf{D}}_{2}$; $R = 0.006\gamma_{0}$, $\gamma = 0.639\gamma_{0}$, $\kappa = (-0.142 + i0.333)\gamma_{0}$ for the H type singlet exciton marked by the cross in Figure 2 (b). (a) Temporal evolution of the Stokes' parameters represented on the Poincar\'e sphere. {Magenta} is for our heterostructure and {green} is for a dipole for which $\text{Im}[\kappa] = 0$ and a similar value of $|\kappa|/\gamma$. We have taken the intervalley scattering rate to be 0 and the system initially to be in one valley. One should note that the system in our heterostructure evolves outside the $xz$ plane. Note that the red dots denote the steady state and are on the $S_{1}-S_{2}$ plane. (b) Steady state Stokes' parameters as a function of the normalized intervalley scattering rate. Since we are considering interlayer excitons, we expect the normalized intervalley scattering rate to be small ($\sim$ 0.1). One can see that the coherences are quite large for such heterostructures. (c) A two-dimensional plot of the phase, $\phi = \atan{(S_{2}/S_{1})}$, marked in (a) for different values of $\gamma_{xx}$ and $\gamma_{yy}$ ($\gamma_{zz} = 1$). The large span of $\phi$ shows a large coverage of the sphere for our dipole.}
\end{figure}
\newpage
\bibliography{Refs.bib}
\end{document}